\documentclass[12pt]{article}
\usepackage{amsmath,amsfonts,amssymb}

\begin{document}
\newcommand{\todo}[1]{{\em \small {#1}}\marginpar{$\Longleftarrow$}}   
\newcommand{\labell}[1]{\label{#1}\qquad_{#1}} 

\rightline{DCPT-08/03}   
\vskip 1cm 

\begin{center} {\Large \bf Asymptotically Plane
    Wave Spacetimes and their Actions}
\end{center} 
\vskip 1cm   
  
\renewcommand{\thefootnote}{\fnsymbol{footnote}} 
\centerline{\bf Julian Le Witt\footnote{j.a.le-witt@durham.ac.uk} and
  Simon F. Ross\footnote{S.F.Ross@durham.ac.uk} }
\vskip .5cm 
\centerline{\it Centre for Particle Theory, Department of Mathematical
  Sciences} 
\centerline{\it Durham University, South Road, Durham DH1 3LE, U.K.}

\setcounter{footnote}{0}   
\renewcommand{\thefootnote}{\arabic{footnote}}   
 
\begin{abstract}
  We propose a definition of asymptotically plane wave spacetimes in
  vacuum gravity in terms of the asymptotic falloff of the metric, and
  discuss the relation to previously constructed exact solutions. We
  construct a well-behaved action principle for such spacetimes, using
  the formalism developed by Mann and Marolf. We show that this action
  is finite on-shell and that the variational principle is
  well-defined for solutions of vacuum gravity satisfying our
  asymptotically plane wave falloff conditions.
\end{abstract}

\section{Introduction}

Plane waves are interesting from a variety of different points of view
(see~\cite{Sadri:2003pr} for a review and further references): they
provide a rich class of exact solutions to Einstein's equations, which
describe the neighbourhood of a null geodesic in any geometry by the
Penrose limit. They also include some maximally supersymmetric
solutions of supergravity, on which the string worldsheet theory is
exactly solvable. Since the seminal work of~\cite{Blau:2002dy} on the
Penrose limit of AdS$_5 \times S^5$, the string theory on the
maximally supersymmetric plane wave has also been of intense interest
as an example of holography~\cite{Berenstein:2002jq}. The spectrum of
strings on the plane wave is related to the spectrum of a quantum
mechanical system obtained from the dual CFT on the boundary of the
AdS$_5$ space. This connection provides stringy tests of the AdS/CFT
correspondence, and has significantly deepened our understanding of
this duality.

However, our understanding of holography for the plane wave is still
incomplete: the duality is more indirect than AdS/CFT, since the dual
quantum mechanics is obtained from the theory on the boundary of AdS,
whereas the Penrose limit which gives rise to the plane wave focuses
on a region at the center of AdS. Although a well-defined notion of
the boundary of the maximally supersymmetric plane was obtained by
conformal compactification in~\cite{Berenstein:2002sa}, and this
boundary turns out to be one-dimensional, a direct connection between
the string theory on this plane wave and a theory living in some sense
on its asymptotic boundary has not yet been constructed. As a result,
it has not been possible to extend the results
of~\cite{Berenstein:2002jq} to discuss a holographic duality for
general plane waves.

One approach to deepening our understanding of the duality for plane
waves is to construct asymptotically plane wave spacetimes, and to
look for interpretations of these spacetimes in field theory terms. In
particular, it is clearly interesting to construct asymptotically
plane wave black holes and black strings. The construction of such
solutions has been discussed
in~\cite{Kaloper:1996hr,Hubeny:2002pj,Hubeny:2002nq,Gimon:2003ms,Gimon:2003xk,Hubeny:2003ug}. The
asymptotic structure of plane waves has also been discussed from a
general point of view, using the causal completion of the spacetime,
in~\cite{Marolf:2002ye,Marolf:2003ik,Marolf:2003jf}.

Another interesting recent development for holography was the
construction of a well-behaved action principle for asymptotically
flat spacetimes in~\cite{Mann:2005yr} (see
also~\cite{Mann:2006bd,Astefanesei:2006zd}), which was argued
in~\cite{Marolf:2006bk} to provide an approach to defining a
holographic dual to asymptotically flat space. This was extended to
study holography for linear dilaton spacetimes
in~\cite{Marolf:2007ys}.

Our aim in this paper is to similarly construct an action principle
for asymptotically plane wave spacetimes, in the hope that this will
shed some light on the issue of holography for plane waves. Our
results may also be useful for other investigations of asymptotically
plane wave spacetimes: for example, these methods can be used to
calculate conserved quantities.

To discuss the action for asymptotically plane wave spacetimes, we
first need a suitable notion of what it means for a spacetime to be
asymptotically plane wave. In section~\ref{falloff}, we propose a
definition in terms of a set of falloff conditions on the metric at
large spatial distances in directions orthogonal to the wave. We start
by assuming that the components of the metric with indices along the
spatial directions orthogonal to the wave fall off as
$\mathcal{O}(r^{2-d})$, where $r$ is a radial coordinate and $d$ is
the number of spatial directions orthogonal to the wave, corresponding
to the influence of a localised source being spread over a
$(d-1)$-sphere at large distances. We then need to determine the
behaviour of the components of the metric with indices parallel to the
wave; we use the linearised equations of motion to relate the falloff
conditions of different components, by assuming that all components
make contributions of the same order to each term in the Einstein
equations. This fixes the falloff of the other components of the
metric. We will show that the known solutions which asymptotically
approach a vacuum plane
wave~\cite{Kaloper:1996hr,Hubeny:2002pj,Hubeny:2002nq} satisfy our
falloff conditions.

We only study solutions of the vacuum Einstein equations; it would
clearly be interesting to extend this to include matter, and in
particular to supergravity. We will see that the black string solution
of~\cite{Gimon:2003xk}, which asymptotically approaches a plane wave
solution in supergravity, does not satisfy our falloff conditions. The
extension to include matter may therefore be non-trivial, as in the
AdS case, where the presence of a scalar field can lead to the
existence of more general AdS-invariant boundary conditions for the
metric~\cite{Hertog:2004dr}.

In section~\ref{action}, we show that the definition of the action for
vacuum gravity introduced in~\cite{Mann:2005yr} can be applied to
asymptotically plane wave spacetimes with our falloff conditions
without significant modification. We demonstrate that the on-shell
action is finite, and that the variational principle is
well-defined. This provides confirmation that this is a useful
definition of asymptotically plane wave, and provides another example
where the counter-term approach of~\cite{Mann:2005yr} is useful,
suggesting that this approach to defining the gravitational action
should have a broad applicability.

We will close the paper in section~\ref{concl} with some comments
and remarks. An open problem for the future is to apply this
definition of the action to calculate the conserved quantities for the
asymptotically plane wave spacetimes. 

\section{Asymptotically plane wave falloff conditions}
\label{falloff}

We consider asymptotically plane wave solutions in vacuum gravity. The
plane wave solutions in $d+2$-dimensional vacuum gravity can be
written in Brinkmann coordinates as
\begin{equation} \label{pwave}
  ds^{2(0)} = -2dx^{+}dx^{-}-\mu_{ij}(x^+) x^{i}x^{j}\left(dx^{+}\right)^{2}+\delta_{ij}dx^{i}dx^{j}, 
\end{equation}
where $i,j=1,\ldots, d$, and $\mu_{ij}(x^+)$ are arbitrary functions
subject only to $\delta^{ij} \mu_{ij}(x^+) = 0$, which ensures that
the solution satisfies the vacuum equations of motion. The coordinates
in the plane wave solution split into two coordinates $x^\pm$ along
the direction of the wave, and the spatial coordinates $x^i$ in the
directions orthogonal to the wave. In the spatial directions, we will
use both Cartesian coordinates $x^i$, and polar coordinates $r,
\theta^a$, $a=1,\ldots (d-1)$:
\begin{equation}
\delta_{ij} dx^i dx^j = dr^2 + r^2 \hat h_{ab} d\theta^a d\theta^b,
\end{equation}
where $\hat{h}_{ab}$ is the metric and $\theta^a$ are the coordinates
on the unit $(d-1)$-sphere $S^{d-1}$. 

A general asymptotically plane wave spacetime will have a metric
$g = g^{(0)} + g^{(1)}$,
where $g^{(1)}$ will have some suitable falloff conditions at large
distance. We will focus on studying the falloff conditions at large
radial distance in the directions orthogonal to the wave.  In the
spatial direction that the wave is travelling in, we will consider
either perturbations which are independent of $x^-$, like the wave
itself, or perturbations which fall off at large $x^-$, but we will
not explicitly specify the falloff conditions in this
direction.\footnote{This is similar to the treatment of linear dilaton
  spacetimes in~\cite{Marolf:2007ys}, where the falloffs in the
  directions along the brane were not explicitly treated.}

Considering first metrics which are independent of $x^-$, we specify
the falloff conditions at large $r$ by making two assumptions. First,
we assume that the spatial components (in the above Cartesian
coordinate system)
$g_{ij}^{\left(1\right)}\sim\mathcal{O}\left(r^{2-d}\right)$. These
are the same falloff conditions as for the spatial components of an
asymptotically flat metric in $d+1$ dimensions. This seems appropriate
because we would expect a perturbation which is independent of $x^-$
to correspond to the effect of a source which is extended along the
direction of the wave, but localised in the transverse spatial
directions, so its effect at large $r$ should be diluted by spreading
on the $S^{d-1}$. 

To fix the falloffs of $g_{\pm\pm}$, $g_{\pm i}$, we make a second
assumption, that all components make contributions of the same order
to each term in the Einstein equations.\footnote{We will not attempt
to fully exploit the information in the asymptotic Einstein equations;
we just use them to determine a set of falloff conditions. The
consistency of our falloff conditions with the dynamical equations of
motion is demonstrated by verifying that the solutions we consider in
the next subsection satisfy our falloff conditions.} This is
essentially a genericity assumption, so it should be appropriate for
finding the general falloff conditions on metric components. In vacuum
gravity, the linearised equations of motion are
$R_{\mu\nu}^{\left(1\right)}=0$, where \cite{Wald:1984rg}
\begin{equation} \label{linRicci}
  R_{\mu\nu}^{\left(1\right)}=-\frac{1}{2}g^{\left(0\right)\rho\sigma}\bigtriangledown_{\rho}^{\left(0\right)}\bigtriangledown_{\sigma}^{\left(0\right)}g_{\mu\nu}^{\left(1\right)}-\frac{1}{2}g^{\left(0\right)\rho\sigma}\bigtriangledown_{\mu}^{\left(0\right)}\bigtriangledown_{\nu}^{\left(0\right)}g_{\rho\sigma}^{\left(1\right)}+g^{\left(0\right)\rho\sigma}\bigtriangledown_{\rho}^{\left(0\right)}\bigtriangledown_{(\mu}^{\left(0\right)}g_{\nu)\sigma}^{\left(1\right)}.
\end{equation}
The idea of our assumption is that the cancellations which give
$R_{\mu\nu}^{\left(1\right)}=0$ should generically involve all the
terms in $R_{\mu\nu}^{\left(1\right)}$. The contribution of $g^{(1)}_{ij}$ to
\eqref{linRicci} gives
\begin{equation}
  R_{ij}^{(1)} \sim \mathcal{O}\left(r^{-d}\right), \quad R_{+i}^{(1)}
  \sim  \mathcal{O}\left(r^{1-d}\right), \quad R_{++}^{(1)} \sim \mathcal{O}\left(r^{2-d}\right).
\end{equation}
Because of the assumption that $g^{(1)}_{ij}$ is constant in $x^-$, it
does not make any contribution to $R_{-i}^{(1)}, R_{+-}^{(1)}$ and $
R_{--}^{(1)}$. Assuming the other terms in
$g_{\mu\nu}^{\left(1\right)}$ produce effects at the same order
determines
\begin{equation}
g_{++}^{(1)} \sim \mathcal{O}\left(r^{4-d}\right), \quad g_{+-}^{(1)}
\sim \mathcal{O}\left(r^{2-d}\right), \quad g_{--}^{(1)} \sim
\mathcal{O}\left(r^{-d}\right), 
\end{equation}
\begin{equation}
g_{+i}^{(1)} \sim \mathcal{O}\left(r^{3-d}\right) ,\quad g_{-i}^{(1)} \sim
\mathcal{O}\left(r^{1-d}\right). 
\end{equation}
With these falloffs, all terms also give 
\begin{equation}
  R_{-i}^{(1)} \sim \mathcal{O}\left(r^{-d-1}\right), \quad R_{+-}^{(1)}
  \sim  \mathcal{O}\left(r^{-d}\right), \quad R_{--}^{(1)} \sim \mathcal{O}\left(r^{-d-2}\right).
\end{equation}

The faster falloff conditions required for metric components with an
$x^-$ index arise because $g^{(0)--} \sim r^2$, so terms in a given
component of $R^{(1)}_{ij}$ coming from $g_{--}^{(1)}$ have an extra
factor of $r^2$ compared to terms coming from
$g_{ij}^{(1)}$. Similarly, the less restrictive conditions on
components with an $x^+$ index are due to the vanishing of
$g^{(0)++}$.

If we consider the more general case, allowing the perturbation to
depend on $x^-$, there will be additional terms in
$R_{\mu\nu}^{\left(1\right)}$ involving derivatives
$\partial_-$. These terms will also come with extra powers of $r$
coming from $g^{(0)--}$. As a result, if we think of a general
perturbation as composed of a part which is independent of $x^-$ and a
part which depends on $x^-$, the part which depends on $x^-$ will be
required to fall off more quickly than the constant
part.\footnote{Even without this additional factor, the $x^-$
  dependent parts would be required to fall off faster than the
  constant parts. The situation is analogous to the solution for a
  localized source described in a cylindrical coordinate system, which
  involves
\begin{equation*}
\frac{1}{(r^2+z^2)^{\frac{d-2}{2}}} \approx \frac{1}{r^{d-2}} -
  \frac{(d-2) z^2}{2 r^{d}} + \ldots,
\end{equation*}
so the $z$-dependent term falls off faster than the constant term at
large $r$. The effect of $g^{(0)--}$ is to make these contributions
fall off even more quickly in the plane wave background.} We find
\begin{equation}
\partial_- g_{ij}^{(1)} \sim \mathcal{O}\left(r^{-d}\right),
\quad \partial_- g_{+j}^{(1)} \sim \mathcal{O}\left(r^{1-d}\right),
\quad \partial_- g_{-j}^{(1)} \sim \mathcal{O}\left(r^{-d-1}\right), 
\end{equation}
\begin{equation}
\partial_- g_{++}^{(1)} \sim \mathcal{O}\left(r^{2-d}\right),
\quad \partial_- g_{+-}^{(1)} \sim \mathcal{O}\left(r^{-d}\right),
\quad \partial_- g_{--}^{(1)} \sim \mathcal{O}\left(r^{-d-2}\right), 
\end{equation}
and
\begin{equation} \label{xmdep2a}
\partial_-\partial_- g_{ij}^{(1)} \sim \mathcal{O}\left(r^{-d-2}\right),
\quad \partial_-\partial_- g_{+j}^{(1)} \sim \mathcal{O}\left(r^{-d-1}\right),
\quad \partial_-\partial_- g_{-j}^{(1)} \sim \mathcal{O}\left(r^{-d-3}\right), 
\end{equation}
\begin{equation} \label{xmdep2b}
\partial_-\partial_- g_{++}^{(1)} \sim \mathcal{O}\left(r^{-d}\right),
\quad \partial_-\partial_- g_{+-}^{(1)} \sim \mathcal{O}\left(r^{-d-2}\right),
\quad \partial_-\partial_- g_{--}^{(1)} \sim \mathcal{O}\left(r^{-d-4}\right). 
\end{equation}
We take the above constraints on the asymptotic falloff of the metric
to define a class of asymptotically plane wave spacetimes. 

Not all of these components of the metric carry independent physical
information; by an appropriate diffeomorphism, we can set some of the
components $g_{\mu\nu}^{(1)}$ to zero at large
distance. In~\cite{Marolf:2007ys}, this diffeomorphism freedom was
fixed by choosing a Gaussian normal gauge, in which the components of
$g_{\mu\nu}^{(1)}$ with radial indices are set to zero. In the present
case, because the directions $x^\pm$ are singled out as special, it
seems more convenient to us to choose a gauge in which
\begin{equation}
g_{+-}^{(1)} = g_{--}^{(1)} = g_{-i}^{(1)} = 0. 
\end{equation}
Because of the faster falloff conditions on the $x^-$ components, the
diffeomorphism which sets these components to zero will not modify the
asymptotic falloff of the other components.

\subsection{Comparison to known solutions}

There have been a few papers on exact solutions of the Einstein
equations which asymptotically approach a plane wave. These provide a
useful check of our analysis: if we have an appropriate set of falloff
conditions, they should be satisfied by these solutions. The first
such solution was constructed in~\cite{Kaloper:1996hr,Hubeny:2002pj},
where a Garfinkle-Vachaspati transform was applied to a black string
solution with a nontrivial scalar field to obtain an asymptotically
plane wave black string,
\begin{equation} \label{gv}
ds^2_{str} = -\frac{2}{h(r)} dx^+ dx^- + \frac{ f(r) + r^2 (3 \cos^2 \theta
  -1)}{h(r)} (dx^+)^2 + (k(r) l(r))^2 (dr^2 + r^2 d\Omega_2^2),  
\end{equation}
\begin{equation}
e^{4 \phi} = \frac{k(r) l(r)}{h^2(r)},
\end{equation}
where
\begin{equation}
f(r) = 1+ \frac{Q_1}{r}, \quad h(r) = 1 + \frac{Q_2}{r}, \quad k(r) =
1 + \frac{P_1}{r}, \quad l(r) = 1 + \frac{P_2}{r}. 
\end{equation}
The presence of the scalar $\phi$ means that this is not a vacuum
solution, but it becomes a vacuum solution at large $r$, and it is
easy to check that our boundary conditions are satisfied. The solution
is independent of $x^-$, and it has $g_{+-}^{(1)}$ and
$g_{ij}^{(1)}$ going like $\mathcal{O}\left(r^{-1}\right)$,
$g_{++}^{(1)}$ going like $\mathcal{O}\left(r\right)$, with
the other components of $g_{\mu\nu}^{(1)}$ vanishing. We have written
the string frame solution above but this statement will be true in
either string or Einstein frame. 

This was extended in~\cite{Hubeny:2002nq} to construct a pure vacuum
solution which is asymptotically plane wave, although it is not smooth
in the interior:
\begin{equation}
ds^2 = \frac{1}{H(r)}\left[-2 dx^+ dx^- + f(r) (dx^+)^2 + \frac{H(r)^4}{r^4
  H'(r)^2} (dr^2 + r^2 d\Omega_2^2) \right], 
\end{equation}
where
\begin{equation}
f(r) = 1 + \ln H(r) + \xi_2(x^+) \psi_2(r) (3 \cos^2 \theta -1), 
\end{equation}
\begin{equation}
\psi_2(r) = ( 3r^2 +2 + 3r^{-2})\left[ \alpha_1 + \alpha_2 \ln \left(
    \frac{r-1}{r+1} \right)  \right] + 6 \alpha_2 (r+ r^{-1}),
\end{equation}
\begin{equation}
H(r) = \left(  \frac{r-1}{r+1} \right)^{\frac{2}{\sqrt{3}}},
\end{equation}
and $\alpha_1$, $\alpha_2$ are arbitrary constants and $\xi_2(x^+)$ is
an arbitrary function of $x^+$. Again, it is easy to see that this
satisfies our definition of asymptotically plane wave. The solution
is independent of $x^-$, and it has $g_{+-}^{(1)}$ and
$g_{ij}^{(1)}$ going like $\mathcal{O}\left(r^{-1}\right)$,
$g_{++}^{(1)}$ going like $\mathcal{O}\left(r\right)$, with
the other components of $g_{\mu\nu}^{(1)}$ vanishing. 

In~\cite{Gimon:2003ms}, a solution was obtained by T-duality from a
black hole in a G\"odel universe. This solution reduces to a plane
wave when the black hole mass parameter is set to zero, but it is not
asymptotically plane wave, as it has components $g_{ij}^{(1)}$ going
like $\mathcal{O}\left(r^{0}\right)$ at large $r$, so the sphere is
deformed asymptotically. Thus, it does not satisfy our definition, but
this is unproblematic: we would not regard such a solution as a
candidate for the appellation asymptotically plane wave.

Finally, another solution was obtained in~\cite{Gimon:2003xk} by a
sequence of boosts and dualities known as the null Melvin twist.  This
is a solution in the common Neveu-Schwarz sector of the
ten-dimensional superstring theories, and has
\begin{eqnarray}
ds^2_{str} &=& - \frac{f(r) (1+ \beta^2 r^2)}{k(r)} dt^2 - \frac{2
  \beta^2 r^2 f(r)} {k(r)} dt dy + \left( 1 - \frac{\beta^2 r^2}{k(r)}
\right) dy^2 \nonumber \\ &&+ \frac{dr^2}{f(r)} + r^2 d\Omega_7^2 - \frac{\beta^2 r^4
  (1-f(r))}{4 k(r)} \sigma^2,
\end{eqnarray}
\begin{equation}
e^\phi = \frac{1}{\sqrt{k(r)}},
\end{equation}
and
\begin{equation}
B = \frac{\beta r^2}{2 k(r)} (f(r) dt + dy) \wedge \sigma,
\end{equation}
where
\begin{equation}
f(r) = 1 - \frac{M}{r^6}, \quad k(r) = 1 + \frac{\beta^2 M}{r^4}, 
\end{equation}
and the one-form $\sigma$ is given in terms of Cartesian coordinates
$x^i$ by
\begin{equation}
\frac{r^2 \sigma}{2} = x^1 dx^2 - x^2 dx^1 + x^3 dx^4 - x^4 dx^3 + x^5
dx^6 - x^6 dx^5 + x^7 dx^8 - x^8 dx^7.  
\end{equation}
This solution is not vacuum, even at large distances, but at large $r$
it approaches a plane wave which~\cite{Gimon:2003xk} call
$\mathcal{P}_{10}$, which is the two-form equivalent of an
electromagnetic plane wave. We can then write the metric as $g =
g^{(0)} + g^{(1)}$, where $g^{(0)}$ is the metric of the pure plane
wave $\mathcal{P}_{10}$, which can be obtained by setting $M=0$ in the
above solution. 

This solution lies outside of the scope of our analysis, since it is
not a solution of the vacuum Einstein equations, even
asymptotically. However, we can still observe that this solution does
not satisfy our asymptotic falloff conditions, as $g^{(1)}_{ij} \sim
\mathcal{O}\left(r^{-4}\right)$, so our input assumption that
$g^{(1)}_{ij} \sim \mathcal{O}\left(r^{2-d}\right)$ is not
satisfied. That is, the spatial falloff of the metric is not behaving
as we would expect based on a localised source, which presumably means
that there are source terms coming from the two-form field $B$ which
extend into the asymptotic region, additional to those associated with
the plane wave $\mathcal{P}_{10}$. In addition, the relation between
the different coefficients is not the same as we had: if we define
$x^+ = t+y$, $x^- = t-y$, we will have $g^{(1)}_{+-} \sim
\mathcal{O}\left(r^{-4}\right)$, but $g^{(1)}_{--} \sim
\mathcal{O}\left(r^{-4}\right)$, and not
$\mathcal{O}\left(r^{-6}\right)$ as we might have expected from the
behaviour of $g^{(1)}_{ij}$. It is not clear whether we should regard
this solution as asymptotically plane wave or not; it asymptotically
approaches the plane wave metric $\mathcal{P}_{10}$, but more slowly
than we would expect. In particular, the slow falloff of the spatial
components $g^{(1)}_{ij}$ is likely to make it difficult to define a
finite action principle for such solutions. It would be very
interesting to extend our analysis below to include form fields so
that this case could be directly addressed.

\subsection{Conformal structure}
\label{conf}

We have given a definition of asymptotically plane wave spacetimes
above, focusing on the behaviour of the solution at large $r$. Our
decision to focus on the behaviour at large $r$ is inspired in part by
the previously-known exact solutions, which approach a plane wave only
at large $r$, and by our interest in the construction of an
appropriate action principle, where it is the $r=$ constant boundary
which is expected to be problematic. 

In special cases, however, we could take a different approach, and
define asymptotically plane wave spacetimes in terms of the existence
of a suitable conformal completion. This would be closer in spirit to
the usual treatments of asymptotic flatness. We will not develop this
approach here; we simply want to make some remarks pointing out that
it is really quite different to the approach we are taking.

In~\cite{Berenstein:2002sa}, a conformal completion was constructed
for the maximally supersymmetric plane wave, for which the metric is
\begin{equation}
ds^2 = -2dx^+ dx^- - r^2 (dx^+)^2 + dr^2 + r^2 d\Omega_7^2,
\end{equation}
where $d\Omega_7^2$ denotes the unit metric on $S^7$. The conformal
completion is obtained by making a coordinate transformation to
rewrite this metric as a conformal factor times the metric on the
Einstein static universe,
\begin{equation}
ds^2 = \frac{1}{|e^{i\psi} - \cos \alpha e^{i\beta}|^2} (-d\psi^2 +
d\alpha^2 + \cos^2 \alpha d\beta^2 + \sin^2 \alpha d\Omega_7^2). 
\end{equation}
We thus see that the conformal boundary of this
plane wave lies at $\alpha =0$, $\psi= \beta$, and is a
one-dimensional null line in the Einstein static universe. The
explicit coordinate transformation is
\begin{equation}
r = \frac{\sin \alpha}{2 |e^{i\psi} - \cos \alpha e^{i\beta}|}, 
\end{equation}
\begin{equation}
\tan x^+ = \frac{\sin \psi - \cos \alpha \sin \beta}{\cos \psi - \cos
  \alpha \cos \beta},  
\end{equation}
\begin{equation}
x^- = \frac{1}{2} \left( \frac{\sin \psi + \cos \alpha \sin \beta}{\cos \psi - \cos
  \alpha \cos \beta} - r^2 \tan x^+ \right). 
\end{equation}

The point we want to stress is that when we approach the conformal
boundary $\alpha = 0$, $\psi - \beta = 0$ along a generic direction,
say $\alpha = \gamma (\psi - \beta)$ for some constant $\gamma$, $r$
remains finite. In these generic directions, it is $x^-$ which
diverges. Thus, controlling the behaviour as $r \to \infty$ in a
spacetime which asymptotically approaches this plane wave will give
little information about whether there exists a conformal completion
with (in some suitable sense) ``the same structure'' as for the pure
plane wave. Rather, it is the behaviour at large $x^-$ that one would
have to study in detail to see if a suitable conformal completion
exists. 

Thus, the definition of asymptotically plane wave we have introduced
is different in character from a definition based on conformal
structure. If a definition based on conformal structure could be
developed, it would presumably be suitable for addressing different
questions from those which can be addressed with our definition. We
would also remark that the above analysis suggests that the known
exact solutions, which have a deformation away from the plane wave
which is independent of $x^-$, are unlikely to qualify as
asymptotically plane wave with respect to such a conformal definition
of asymptotically plane wave. 

\section{Action for asymptotically plane wave spacetimes}
\label{action}

We have put forward a definition of asymptotically plane wave
spacetimes, using the linearised equations of motion to relate the
falloff of different components. In this section, we give our main
result, constructing an appropriate action principle for this class of
spacetimes. We construct our action principle following Mann and
Marolf \cite{Mann:2005yr}, who recently introduced a new approach to
specifying a well-defined action principle for vacuum gravity for
asymptotically flat spacetimes. 

For the asymptotically flat case, the action is~\cite{Mann:2005yr}
\begin{equation}
S=-\frac{1}{16\pi G}\int_{M}\sqrt{-g} R d^{D}x-\frac{1}{8\pi
  G}\int_{\partial M}\sqrt{h} K d^{D-1}x+\frac{1}{8\pi G}\int_{\partial
  M}\sqrt{h} \hat K d^{D-1}x,
\end{equation}
where $g$ is the determinant of the bulk metric, $h$ is the
determinant of the bulk metric pulled back to the boundary, $R$ is the
Ricci scalar, and $K=h^{\alpha\beta}K_{\alpha\beta}$ is the trace of
the extrinsic curvature on the boundary. The final term is a new
contribution introduced to cancel the divergences coming from the
Gibbons-Hawking boundary term (the second term above). The function
$\hat{K}$ is defined implicitly by the solution of
\begin{equation}\label{khat}
  R_{\alpha\beta}=\hat{K}{_{\alpha\beta}}\hat{K}-h^{\gamma\delta}\hat{K}{_{\alpha\gamma}}\hat{K}{_{\delta\beta}},
\end{equation} 
where $R_{\alpha\beta}$ is the Ricci tensor of the metric
$h_{\alpha\beta}$ induced on $\partial M$. Thus this additional
boundary term is determined locally by the induced metric on the
boundary, in the spirit of the boundary counterterm approach to
constructing actions for asymptotically AdS
spaces~\cite{Balasubramanian:1999re}. Alternative actions for 
asymptotically flat spacetimes with a similar philosophy appeared
previously in~\cite{Mann:1999pc,Kraus:1999di}. See also
\cite{Cai:1999xg} for related work.

To apply this prescription to asymptotically plane wave spacetimes, we
first need to introduce a cutoff to make the different terms in the
action finite. We will cut off the spacetime by introducing a boundary
at some large radial distance, $r =$ constant. Our main focus will
be on boundary terms associated with this boundary; as in the
asymptotically flat case, there is a divergence associated with the
Gibbons-Hawking boundary term on this surface due to the extrinsic
curvature of the sphere, and we need to introduce an appropriate local
boundary term to cancel it. 

Although our focus is mainly on the $r=$ constant boundary, to make
the spacetime region we consider finite, we also need to introduce
some cutoffs in the $x^\pm$ directions along the plane wave. The
symmetry of the background under translations in $x^-$ makes it
natural to introduce cutoffs at two constant values of $x^+$,
respecting this symmetry. In the simple case where $\mu_{ij}$ are
constants, which includes the cases of most interest for holography,
there is an additional symmetry under translations in $x^+$, which
suggests it is natural to take the other cutoff to be at constant
values of $x^-$, respecting this translation invariance. We will also
discuss the calculation of the action for the general case where
$\mu_{ij}(x^+)$ are not constants with this same cutoff. We will see
that this choice of cutoff can give a satisfactory construction for an
action even for general $\mu_{ij}(x^+)$, although there are some
additional subtleties associated with the surfaces at constant
$x^-$. However, one should bear in mind that there is no a priori
justification for this choice of cutoff in the general case.

The action for the cutoff spacetime should contain a Gibbons-Hawking
boundary term for each of these boundaries. In the case of the surfaces
at $x^+ = $ constant, there is a subtlety, as they are null surfaces,
so the trace of the extrinsic curvature is not well-defined. However,
this issue has been previously considered in~\cite{Barrabes:2005ar},
where it was shown that a suitable boundary term on a null boundary
$x^+ = $ constant is
\begin{equation}
\frac{1}{16 \pi G}
\int_{x^{+}=const}d^{d+1} \xi \sigma^{\lambda}\partial_{\lambda}x^{+} ,
\end{equation}
where
$\sigma^{\lambda}=\frac{1}{\sqrt{-g}}\partial_{\mu}\left(\left(-g\right)g^{\mu\lambda}\right)$,
with $g$ being the determinant of the metric on the full spacetime. We
will adopt this prescription here. On the boundaries at $x^- = $
constant, we consider just the usual Gibbons-Hawking boundary term.

On the boundary at $r=$ constant, the Gibbons-Hawking boundary term
gives a contribution which will diverge as we remove the cutoff. This
divergence is associated with the intrinsic curvature of the boundary
(the background plane wave spacetime has a flat spatial metric in the
$x^i$ directions, so the intrinsic and extrinsic curvatures of the
$r=$ constant boundary are related), so we can try to cancel this
divergence by adding a Mann-Marolf counterterm contribution to the
action on this boundary.

Thus, the action we consider is 
\begin{eqnarray}
S&=&-\frac{1}{16\pi G}\int_{M}d^{d+2}x \sqrt{-g}  R  - \frac{1}{16 \pi G}
\int_{x^{+}=consts}d^{d+1}x \sigma^{\lambda}\partial_{\lambda}x^{+}
\nonumber \\ &&-\frac{1}{8\pi
  G}\int_{x^{-}=consts} d^{d+1}x \sqrt{h} K -\frac{1}{8\pi
  G}\int_{r=const}d^{d+1}x \sqrt{h} \left(K-\hat{K}\right),
\end{eqnarray}
where by the integral over $x^+ =$ constants we mean integrals over
two surfaces at different values of $x^+$, with opposite orientations
for the normal to the surface, and similarly for the integral over
$x^- =$ constants.

Let us first of all consider the value of this action for the plane
wave background~\eqref{pwave}. This is a vacuum solution, so $R=0$. On
the surface $x^+ =$ constant, 
\begin{equation}
\sigma^\lambda \partial_\lambda x^+ =
\sigma^+ = \partial_\mu g^{\mu+} = 0,
\end{equation}
as $g^{(0)++} = 0$ and $g^{(0)+-} = -1$. So the boundary term at
$x^+=$ constant vanishes. On the surface $x^-=$ constant, if
$\mu_{ij}$ are constant, the only non-zero component of $K_{\alpha
  \beta}$ is
\begin{equation}
K_{+i} =
  \frac{1}{2\sqrt{g^{(0)--}}} \partial_i g^{(0)}_{++}. 
\end{equation}
Since $h^{(0)+i}=0$, this gives $K=0$, and the boundary term at
$x^-=$ constant vanishes as well.

In the more general case where $\mu_{ij}(x^+)$ depend on $x^+$, we have
\begin{equation}
K = K_{++} h^{(0)++} = \frac{1}{2 \sqrt{g^{(0)--}}} \partial_+
  g^{(0)}_{++} h^{(0)++},
\end{equation}
and at $x^-=$ constant, $h^{(0)++} = 1/h^{(0)}_{++} =
-1/(\mu_{ij}(x^+) x^i x^j)$. Hence, this $K \sim \mathcal O(r^{-1})$,
and the contribution to the action is 
\begin{equation}
S_- = -\frac{1}{8\pi G} \int_{x^- = const} K \sqrt{h} dx^+ d^{d} x^i
\sim \mathcal O(r^d),  
\end{equation}
so this boundary will make a divergent contribution to the action as
we remove the cutoff at large $r$. However, in the full action, there
are two boundaries at constant $x^-$ (at say $x^- = \pm x^-_0$), and
they contribute with opposite signs because of the opposite
orientations of the outward normals, so this term will cancel between
the two boundaries, making no contribution to the total action. 

Finally, the boundary at $r=$ constant is what we want to focus on, so
let us be more explicit and set up the notation we will use
later. Define coordinates on the boundary $x^{\alpha}=\left\{
  x^{-},x^{+},\theta^{a}\right\} $, so the boundary metric is
\begin{equation}
h_{\alpha\beta}=\left(\begin{array}{ccc}
0 & -1 & \vec{0}\\
-1 & -\mu_{ij}x^{i}x^{j} & \vec{0}\\
\vec{0} & \vec{0} & r^{2}\hat{h}{}_{ab}\end{array}\right),
\end{equation}
with determinant $h = -r^{2d-2} \hat h$, where $\hat{h}$ is the
determinant of the unit metric on $S^{d-1}$.  The normal vector to the
boundary is $n_{\nu}=\delta_{\nu}^{r}$.  The non-zero components of
the extrinsic curvature are
\begin{equation}
K_{ab}=r\hat{h}_{ab}, \quad K_{++}=-\frac{\mu_{ij}x^{i}x^{j}}{r},
\end{equation}
so $K=\frac{d-1}{r}$.  The Ricci tensor on the boundary is
\begin{equation}
R_{\alpha\beta}=\left(\begin{array}{ccc}
0 & 0 & \vec{0}\\
0 & R_{++} & \vec{0}\\
\vec{0} & \vec{0} &
\left(d-2\right)\hat{h}{}_{ab}\end{array}\right).
\end{equation}
Solving \eqref{khat} for $\hat{K}{_{\alpha\beta}}$, we find that the
non-zero components are 
$\hat{K}{_{ab}}=r\hat{h}{}_{ab}$ and $\hat{K}{_{++}}=\frac{rR_{++}}{d-1},$
and so $\hat{K}{}=\frac{d-1}{r}$. Thus $K-\hat{K}=0$ and hence there
is no contribution to the action from the $r=$ constant surface. 

Thus, we find that the on-shell action for the pure plane wave is
zero. Note that the action vanishes for any plane wave solution,
independent of the values of $\mu_{ij}(x^+)$.

\subsection{Finiteness of the action}

Next, we consider an arbitrary asymptotically plane wave solution
satisfying our asymptotic falloff conditions, and show that the action
of the solution will be finite. Since the metric $g$ is still a
solution of the vacuum equations, $R=0$, so the bulk term still makes
no contribution to the action. For the boundaries at constant $x^+$,
as in the pure plane wave,
\begin{equation}
S_{+}=-\frac{1}{16 \pi G} \int_{x^{+}=const}dx^{-}\left(dx^{i}\right)^{d}\partial_{\mu}g^{(1)\mu+}.
\end{equation}
In the gauge we have chosen, $g^{++} = 0$, $g^{+-}=1$, and $g^{+i}=0$,
so this term still vanishes. 

For the boundaries at constant $x^-$, the contributions to the
extrinsic curvature at linear order in the departure of the metric
from the plane wave are
\begin{equation}
  K = K_{++}^{(0)} h^{(1)++}+ K_{+i}^{(0)} h^{(1)+i} + K_{++}^{(1)}h^{(0)++}+ K_{ij}^{(1)}h^{(0)ij}.
\end{equation}
On these boundaries, we have $h^{(1)++} \sim \mathcal O(r^{-d})$,
$h^{(1)+i} \sim \mathcal O(r^{1-d})$, and 
\begin{equation}
K_{++}^{(1)} =
-\frac{1}{2}\frac{g^{(0)+-}}{\sqrt{g^{(0)--}}} \partial_+ g^{(1)}_{++}-
\frac{1}{2}\frac{g^{(0)+-} g^{(1)--}}{(g^{(0)--})^{3/2}} \partial_+ g^{(0)}_{++}
+\frac{1}{2}\sqrt{g^{(0)--}} \partial_- g^{(1)}_{++},
\end{equation}
\begin{equation}
K_{ij}^{(1)} =-\frac{1}{2}\frac{g^{(0)+-}}{\sqrt{g^{(0)--}}}\left(\partial_j
   g_{i+}^{(1)}+\partial_i g_{j+}^{(1)}-\partial_+
   g_{ij}^{(1)}\right) +\frac{1}{2}\sqrt{g^{(0)--}}\partial_-
   g_{ij}^{(1)} . 
\end{equation}
Thus, the terms which are independent of $x^-$ will give a
contribution to $K \sim \mathcal O(r^{1-d})$. This will make a
divergent contribution to the integral over a single boundary, $S_-
\sim \mathcal O(r^2)$. However, as in the action for the pure plane
wave, this divergence cancels between the two boundaries, so for
asymptotically plane wave solutions which are independent of $x^-$,
the contribution to the action from these boundaries vanishes.

We require that any terms depending on $x^-$ fall off at large
$x^-$. This implies in particular that there cannot be any linear
dependence on $x^-$ near these boundaries, so the part of the
components $g^{(1)}_{\mu\nu}$ involving $x^-$ will fall off faster
than the part that is independent of $x^-$ by a factor of $1/r^4$. The
contribution of the $x^-$-dependent part of $g^{(1)}_{\mu\nu}$ to the
terms in $K$ that do not involve explicit derivatives $\partial_-$
will then be $O(r^{-d-3})$. Thus the contribution to the action from
this part of $K$ is finite, and will go to zero as we take the cutoff
in $x^-$ to infinity. There are terms in $K_{++}^{(1)}$ and
$K_{ij}^{(1)}$ which involve explicit derivatives $\partial_-$: these
make a contribution $K \sim \mathcal O(r^{-d-1})$, giving a
contribution to the integral $S_-$ which is logarithmically divergent
at large $r$. However, this contribution comes with some negative
power of $x^-$, so if we take the boundaries at constant $x^-$ to
infinity at the same time as we take the boundary at large $r$ to
infinity, this contribution will go to zero. This dependence on the
order of limits is not entirely satisfactory, but it allows us to
define a finite action. It does not seem to conceal any particularly
interesting deeper issues. 

Finally, we consider the boundary at $r=$ constant. We can write the
linear order contribution to the boundary term in our gauge as
\begin{equation}
K^{(1)}-\hat{K}^{(1)}=K_{\alpha\beta}^{(1)}h^{(0)\alpha\beta}-\hat{K}{_{\alpha\beta}^{(1)}}h^{(0)\alpha\beta}. 
\end{equation}
As $\sqrt{h} \sim \mathcal O(r^{d-1})$, we need $K^{(1)}-\hat{K}^{(1)}
\sim \mathcal O(r^{1-d})$ to have a finite action.  For the term
involving the extrinsic curvature,
\begin{equation}
K_{\alpha\beta}^{(1)}=g^{(1)rr}K_{\alpha\beta}^{(0)}+\frac{1}{2}\left(g_{\beta
    r,\alpha}^{(1)}+g_{r\alpha,\beta}^{(1)}-g_{\alpha\beta,r}^{(1)}\right),
\end{equation}
and substituting for $g_{\alpha\beta}^{(1)}$ it is easy to show that
this term is $\mathcal O(r^{1-d})$.

To evaluate $\hat{K}{_{\alpha\beta}^{(1)}}$, we linearize \eqref{khat}
to give
\begin{equation} \label{linR}
R_{\alpha\beta}^{(1)}=\hat{K}{_{\gamma\delta}^{(1)}}L_{\alpha\beta}^{(0)\gamma\delta}+\left(\hat{K}{_{\alpha\beta}^{(0)}}\hat{K}{_{\gamma\delta}^{(0)}}-\hat{K}{_{\alpha\gamma}^{(0)}}\hat{K}{_{\beta\delta}^{(0)}}\right)h^{(1)\gamma\delta},
\end{equation}
where\footnote{Note that we define $L_{\alpha\beta}^{(0)\gamma\delta}$ so that it is
symmetric in both pairs of indices, so this is slightly different from
the corresponding expression in~\cite{Marolf:2007ys}. }
\begin{equation}
L_{\alpha\beta}^{(0)\gamma\delta}=h^{\gamma\delta}\hat{K}{_{\alpha\beta}}+\frac{1}{2}\left(\delta_{\alpha}^{\gamma}\delta_{\beta}^{\delta}\hat{K}+\delta_{\beta}^{\gamma}\delta_{\alpha}^{\delta}\hat{K}\right)-\frac{1}{2}\left(\delta_{\alpha}^{\gamma}\hat{K}{_{\beta}^{\delta}}+\delta_{\beta}^{\gamma}\hat{K}{_{\alpha}^{\delta}}+\delta_{\alpha}^{\delta}\hat{K}{_{\beta}^{\gamma}}+\delta_{\beta}^{\delta}\hat{K}{_{\alpha}^{\gamma}}\right).
\end{equation}
Inverting this
will give us an expression for $\hat{K}{_{\alpha\beta}^{(1)}}$, 
\begin{equation}
h^{(0)\alpha\beta}\hat{K}{_{\alpha\beta}^{(1)}}=M^{(0)\gamma\delta}\left(R_{\gamma\delta}^{(1)}-\left(\hat{K}{_{\alpha\beta}^{(0)}}\hat{K}{_{\gamma\delta}^{(0)}}-\hat{K}{_{\alpha\gamma}^{(0)}}\hat{K}{_{\beta\delta}^{(0)}}\right)h^{(1)\alpha\beta}\right),
\end{equation}
where
$M^{\gamma\delta}=h^{\alpha\beta}\left(L^{-1}\right)_{\alpha\beta}^{\
  \ \gamma\delta}$. Recall that the non-zero components in
$\hat{K}{_{\alpha\beta}^{(0)}}$ are $\hat{K}{_{++}^{(0)}}$ and
$\hat{K}{_{ab}^{(0)}}$, and note that in our gauge $h^{(1)++} =0$ on
the $r=$ constant boundary. We thus have
\begin{equation} \label{k1hatr}
  h^{(0)\alpha\beta}\hat{K}{_{\alpha\beta}^{(1)}}=M^{(0)\alpha \beta
  }R_{\alpha
    \beta}^{(1)}-M^{(0)ab}(\hat{K}{_{ab}^{(0)}}\hat{K}{_{cd}^{(0)}}-\hat{K}{_{ac}^{(0)}}\hat{K}{_{bd}^{(0)}})h^{(1)cd}.
\end{equation}
A lengthy explicit calculation gives that the only non-zero components
of $M^{(0)\gamma\delta}$ are 
\begin{equation}
M^{(0)+-} \sim \mathcal O(r), \quad M^{(0)--} \sim \mathcal O(r^2),
\quad M^{(0)ab} = \frac{1}{2(d-2)r} \hat{h}^{ab} = \frac{r}{2(d-2)} h^{ab}.
\end{equation}
For the second term in \eqref{k1hatr}, we have $\hat{K}{_{ab}^{(0)}}
\sim \mathcal O(r)$, and $h^{(1)cd} \sim \mathcal O(r^{-d})$, so this
term is $\mathcal O(r^{1-d})$. For the first term, we express
$R_{\alpha\beta}^{(1)}$ by the analogue of \eqref{linRicci},
\begin{equation}
R_{\alpha\beta}^{(1)}=-\frac{1}{2}h^{(0)\gamma\delta}D_{\alpha}^{(0)}D_{\beta}^{(0)}h_{\gamma\delta}^{(1)}-\frac{1}{2}h^{(0)\gamma\delta}D_{\gamma}^{(0)}D_{\delta}^{(0)}h_{\alpha\beta}^{(1)}+h^{(0)\gamma\delta}D_{\gamma}^{(0)}D_{(\alpha}^{(0)}h_{\beta)\delta}^{(1)},
\end{equation}
where $D_{\alpha}$ is the covariant derivative compatible with
$h_{\alpha\beta}$. Using this expression we can see that $R_{+-}^{(1)}
\sim \mathcal O(r^{-d})$, $R_{--}^{(1)} \sim \mathcal O(r^{-d-2})$,
and $R_{ab}^{(1)} \sim \mathcal O(r^{2-d})$, so the first term also
makes a finite contribution (in addition, many of these terms will
actually be total derivatives, which make no contribution to the
action). 

Thus, we conclude that the on-shell action is finite for the
asymptotically plane wave spacetimes.

\subsection{Variations of the action}

In addition to being finite on-shell, we would like to see that
$\delta S=0$ for arbitrary variations about a solution of the
equations of motion. The variation of the usual Einstein-Hilbert plus
Gibbons-Hawking action would give a boundary term 
\begin{equation}
\delta S_{EH+GH}=-\frac{1}{16\pi G}\int\sqrt{-h}\pi^{\alpha\beta}\delta
h_{\alpha\beta},
\end{equation}
where $\pi^{\alpha \beta} = K^{\alpha \beta} - h^{\alpha \beta} K$. On
the boundaries at $x^+ = $ constant and $x^- =$ constant, we have just
this term. Therefore if we require $\delta h_{\alpha \beta}=0$ on
these boundaries, they will make no contribution to the variation of
the action. This is a reasonable boundary condition if we think of
these as fixed cutoffs; that is, if we will keep the coordinate
position of the cutoff fixed as we vary the metric, and do not intend
to eventually send the cutoff to infinity. This is certainly an
appropriate approach for the $x^+ =$ constant boundary. In some cases,
however, it is more appropriate to eventually remove the cutoff on
$x^-$. For this purpose, we could imagine relaxing this condition to
require only that $\delta h_{\alpha \beta}$ decays as we go to large
$x^-$. Since the background metric is independent of $x^-$, any
$\delta h_{\alpha \beta}$ which goes to zero at large $x^-$ will
produce a contribution to $\delta S$ which vanishes as we remove the
cutoff on $x^-$. Thus, there is no problem with the variation of the
action involving these boundaries.

We turn to the contribution to the variation of the action from the
boundary at $r=$ constant, where we only want to require that the
variation $\delta h_{\alpha \beta}$ falls off as quickly as
$g^{(1)}_{\alpha \beta}$. On the $r= $ constant boundary, we have the
above boundary contribution from the Einstein-Hilbert plus
Gibbons-Hawking action, and we have the contribution coming from the
variation of the new boundary term,
\begin{equation}
\delta S_{MM}=\frac{1}{8\pi
  G}\int\sqrt{-h}\left(-\frac{1}{2}\hat{K}h^{\alpha\beta}\delta
  h_{\alpha\beta}+\hat{K}{_{\alpha\beta}}\delta
  h^{\alpha\beta}+h^{\alpha\beta}\delta\hat{K}{_{\alpha\beta}}\right).
\end{equation}
To determine $h^{\alpha\beta}\delta\hat{K}{_{\alpha\beta}}$, we need
to use the analogue of \eqref{linR} for variations to write
\begin{equation} \label{varK}
h^{\alpha\beta}\delta\hat{K}{_{\alpha\beta}} = M^{\gamma\delta}\left(\delta
R_{\gamma\delta}-\left(\hat{K}_{\alpha\beta}\hat{K}_{\gamma\delta}-\hat{K}_{\alpha\gamma}\hat{K}_{\beta\delta}\right)\delta
h^{\alpha\beta}\right),
\end{equation}
where $\delta R_{\gamma \delta}$ is given in terms of $\delta
h_{\alpha \beta}$ by 
\begin{equation} \label{varR}
\delta R_{\alpha\beta}=-\frac{1}{2}h^{\gamma\delta}D_{\alpha}D_{\beta}\delta
h_{\gamma\delta}-\frac{1}{2}h^{\gamma\delta}D_{\gamma}D_{\delta}\delta
h_{\alpha\beta}+h^{\gamma\delta}D_{\gamma}D_{(\alpha}\delta
h_{\beta)\delta}. 
\end{equation}
The variation can be taken to respect our choice of gauge, so $\delta
h_{-\alpha}=0$. Thus, we only need to consider the variations $\delta
h_{++}$, $\delta h_{+a}$ and $\delta h_{ab}$. 

Let's consider first just $\delta h_{++}$ non-zero. The term in
$\delta S_{EH+GH}$ involving $\delta h_{++}$ is trivially zero, as
$\pi^{++} = 0$ with our choice of gauge. For the new boundary term,
\begin{equation} \label{hppvar}
\delta
S_{MM}=\frac{1}{8\pi G}\int\sqrt{-h}\left(\hat{K}{^{++}}\delta
  h_{++}+h^{\alpha\beta}\delta\hat{K}{_{\alpha\beta}}\right).
\end{equation}
This expression involves the full metric of the asymptotically plane
wave solution we are considering. For each term, we will explicitly
calculate the result for the leading non-zero contribution (coming
either from $g^{(0)}$ or $g^{(1)}$). Higher-order terms are
suppressed, so if the first term gives zero contribution to the
variation of the action, we do not need to consider higher orders.  In
the first term in \eqref{hppvar}, solving for $\hat{K}{^{(1)++}}$
using \eqref{linR} gives $\hat{K}{^{(1)++}} \sim \mathcal
O(r^{-d-1})$, and $\delta h_{++} \sim \mathcal O(r^{4-d})$, so
$\hat{K}^{++} \delta h_{++} \sim \mathcal O(r^{3-2d})$, and the first
term in the integral is $\mathcal O(r^{2-d})$, which vanishes for $d
\geq 3$. For the second term, we use \eqref{varK}, where there will be
a zeroth-order contribution to the first term and a first-order
contribution to the second term. From \eqref{varR}, we find that
$\delta h_{++}$ gives only $\delta R_{++}$, $\delta R_{+-}$ and
$\delta R_{+a}$ non-zero. Using our previous calculation of the
components $M^{(0)\alpha \beta}$, we then have
\begin{equation}
h^{\alpha\beta}\delta\hat{K}{_{\alpha\beta}}= M^{(0)+-} \delta R^{(0)}_{+-}-
M^{(0)ab} \hat{K}^{(0)}_{ab} \hat{K}{^{(1)++}}\delta h_{++}.
\end{equation}
Now $\delta R^{(0)}_{+-} = -\frac{1}{2} h^{(0)+-} \partial_-
\partial_- \delta h_{++} \sim \mathcal O(r^{-d})$, so the first term
is $\mathcal O (r^{1-d})$. Together with the factor of $\sqrt{-h}$ in
the integral, this would give a finite contribution to the
variation. However, this leading-order term is a total derivative,
because $h^{(0)}_{\alpha \beta}$ is independent of $x^-$, so it makes
no contribution. Higher-order contributions from this term would not
be a total derivative, but they are suppressed by further powers of
$r$, so their contribution to the action vanishes in the large $r$
limit. The second term is of the same form as the contribution
considered above, giving a contribution $h^{\alpha \beta} \hat
K_{\alpha \beta} \sim \mathcal O(r^{3-2d})$. Thus all the terms coming
from $\delta h_{++}$ vanish in the large $r$ limit.

We now evaluate terms involving $\delta h_{a+}$. We find 
\begin{equation}
\delta S_{EH+GH}=-\frac{1}{16\pi G}\int\sqrt{-h} \pi^{a+} \delta
h_{a+}. 
\end{equation}
At linear order, $\pi^{a+} \sim h^{ab} \partial_- h_{br} \sim \mathcal
O(r^{-d-1})$, and $\delta h_{a+} \sim \mathcal O(r^{4-d})$, so this
term is vanishing for $d \geq 3$. For the new boundary term, 
\begin{equation}
\delta S_{MM}=\frac{1}{8\pi G}\int\sqrt{-h}\left(\hat{K}{^{a+}}\delta
  h_{a+}+h^{\alpha\beta}\delta\hat{K}{_{\alpha\beta}}\right), 
\end{equation}
and \eqref{linR} gives $\hat{K}^{(1)a+} \sim \mathcal O(r^{-d-1})$, so
the first term also vanishes for $d \geq 3$. In the second term,
having just $\delta h_{a+}$ gives us all components of $\delta
R_{\alpha \beta}$ except $\delta R_{--}$ non-zero. Using \eqref{varK}
and our previous calculation of the components $M^{(0)\alpha \beta}$,
we then have
\begin{equation}
h^{\alpha\beta}\delta\hat{K}{_{\alpha\beta}}= M^{(0)+-} \delta R^{(0)}_{+-}+
M^{(0)ab} \delta R^{(0)}_{ab}- M^{(0)ab} \hat{K}^{(0)}_{ab} \hat{K}^{(1)a+}\delta h_{a+}.
\end{equation}
We have $\delta R^{(0)}_{+-} = \frac{1}{2} h^{(0)cb}
D^{(0)}_b \partial_- \delta h_{+c} \sim \mathcal O(r^{-d})$, and
$\delta R^{(0)}_{ab} =  \frac{1}{2} h^{(0)+-}
\partial_- D^{(0)}_b \delta h_{a+} \sim \mathcal O(r^{2-d})$. Thus,
both of the first two terms in
$h^{\alpha\beta}\delta\hat{K}{_{\alpha\beta}}$ would make finite
contributions to the variation of the action. However, as they involve
$\partial_-$, they are total derivatives, so they actually make zero
contribution. As in the previous case when we analysed terms involving
$\delta h_{++}$, higher-order contributions from this term would not
be a total derivative, but they are suppressed by further powers of
$r$, so their contribution to the action vanishes in the large $r$
limit.  The final term in
$h^{\alpha\beta}\delta\hat{K}{_{\alpha\beta}}$ is of the same form as
the contribution to the variation coming from $\hat K^{a+} \delta
h_{a+}$, so it goes like $\mathcal O(r^{3-2d})$, and all the terms in
the variation of the action coming from $\delta h_{a+}$ vanish in the
large $r$ limit.

Finally we consider terms involving $\delta h_{ab}$. We find 
\begin{equation} \label{ehrvar}
\delta S_{EH+GH}=-\frac{1}{16\pi G}\int\sqrt{-h}\pi^{ab}\delta h_{ab},
\end{equation}
and since $\pi^{ab} \sim \mathcal O(r^{-3})$ and $\delta h_{ab} \sim
\mathcal O(r^{4-d})$, this gives an $r^0$ term which does not vanish
in the large $r$ limit. This term needs to be cancelled by a
corresponding term coming from $\delta S_{MM}$. The latter is 
\begin{eqnarray}
\delta S_{MM} &=& \frac{1}{8\pi G} \int\sqrt{-h}\left( - \frac{1}{2}
  \hat{K} h^{\alpha \beta}\delta h_{\alpha \beta} + \hat K_{\alpha
    \beta} \delta h^{\alpha \beta}
  +h^{\alpha\beta}\delta\hat{K}{_{\alpha\beta}}\right) \nonumber \\
&=&\frac{1}{8\pi
  G}\int\sqrt{-h}\left(\frac{1}{2}\hat{\pi}{^{ab}}\delta
  h_{ab}+\frac{1}{2}\hat{K}{^{ab}}\delta
  h_{ab}+h^{\alpha\beta}\delta\hat{K}{_{\alpha\beta}}\right), \label{mmrvar}
\end{eqnarray}
where $\hat{\pi}{^{ab}}=\hat{K}{^{ab}}-h^{ab}\hat{K}$. To zeroth
order, $\hat \pi^{(0)ab} = \pi^{(0)ab}$, so the first term
in \eqref{mmrvar} cancels the non-zero contribution from
\eqref{ehrvar}. However, the second term in \eqref{mmrvar} also has a
non-zero leading order part, so we need to see that this can be
cancelled by a contribution from the final term. Considering the
variation $\delta h_{ab}$, 
\begin{eqnarray} \label{khatrvar}
h^{\alpha\beta}\delta\hat{K}{_{\alpha\beta}}&=& M^{(0)+-} \delta
R^{(0)}_{+-}+ M^{(0)--} \delta R^{(0)}_{--} + M^{(0)ab} \delta R^{(0)}_{ab}
\\ &&- M^{(0)ab} \left(\hat{K}^{(0)}_{ab}\hat{K}^{(0)}_{cd}-\hat{K}^{(0)}_{ac}\hat{K}^{(0)}_{bd}\right)\delta
h^{cd}. \nonumber  
\end{eqnarray}
The terms involving $\delta R_{\alpha \beta}$ give finite
contributions which are total derivatives, as before. For the first
two terms, 
\begin{equation}
\delta R^{(0)}_{+-} = h^{(0)ab} D^{(0)}_+ \partial_- \delta h_{ab}
\sim \mathcal O(r^{-d}), \quad \delta R^{(0)}_{--} =
h^{(0)ab} \partial_- \partial_- \delta h_{ab} \sim \mathcal O(r^{2-d}),
\end{equation}
and these are total derivatives because $h^{(0)}_{\alpha \beta}$ is
independent of $x^-$. For the other term, $\delta R^{(0)}_{ab} \sim
\mathcal O(r^{2-d})$ involves covariant derivatives with respect to
the unit metric on $S^{d-1}$, $\hat h_{ab}$, and this term is a total
derivative because the only $\theta_a$ dependence in the terms
multiplying $\delta R^{(0)}_{ab}$ is through the covariantly constant
metric $\hat h_{ab}$. As in the previous two cases, higher-order
contributions from these terms would not be total derivatives, but
they are suppressed by further powers of $r$, so their contribution to
the action vanishes in the large $r$ limit. We are then left with
evaluating the last term in \eqref{khatrvar}. Using
$\hat{K}^{(0)}_{ab} = r \hat{h}_{ab}$ and $M^{(0)ab} =
\frac{1}{2(d-2)} \hat{h}^{ab}$, 
\begin{equation}
h^{\alpha\beta}\delta\hat{K}{_{\alpha\beta}} \rightarrow - M^{(0)ab} \left(\hat{K}^{(0)}_{ab}\hat{K}^{(0)}_{cd}-\hat{K}^{(0)}_{ac}\hat{K}^{(0)}_{bd}\right)\delta
h^{cd} = -\frac{1}{2}r
\hat{h}^{ab} \delta h_{ab} = -\frac{1}{2} \hat{K}^{(0)ab} \delta
h_{ab}.
\end{equation}
This will cancel with the leading order part of the second term in
\eqref{mmrvar}, leaving us with no finite contributions to the
variation of the action in the large $r$ limit.

Thus, this action gives a well-defined variational principle for our
class of asymptotically plane wave spacetimes.

\section{Conclusions}
\label{concl}

In this paper, we have given a definition of asymptotically plane wave
spacetimes which is consistent with the known exact solutions, and
constructed a well-behaved action principle for asymptotically plane
wave solutions of the vacuum Einstein equations, following the work
of~\cite{Mann:2005yr}. The definition of asymptotically plane wave
solutions is valid for any solution which asymptotically approaches a
vacuum plane wave. For the action, we considered only the pure vacuum
action; it would be interesting to extend this work to include
appropriate matter fields. It is also interesting to ask if there are
non-trivial physically relevant examples to which our ideas
apply.\footnote{We thank the referee for raising this point.} For the
asymptotically plane wave boundary conditions, \eqref{gv} provides
such an example, but this is not a pure vacuum solution, so our
discussion of the action does not apply to it. A more trivial example
is provided by some pp-wave solutions. For example, consider the
vacuum pp-wave metric
\begin{equation} \label{ppwave}
  ds^{2} = -2dx^{+}dx^{-}-A(x^+,x^i)
  \left(dx^{+}\right)^{2}+\delta_{ij}dx^{i}dx^{j} 
\end{equation}
with $\partial_i \partial^i A = 0$. If $A(x^+, x^i) \to \mu_{ij}(x^+)
x^i x^j + \mathcal O(r^{4-d})$ as $r \to \infty$, this solution is
asymptotically plane wave according to our definition, and the action
we have defined will be finite for it. However, this is a rather
trivial example, and it would be interesting to construct solutions
really corresponding to localised sources in an asymptotically plane
wave background, and we hope to return to this question in future work.

We have just demonstrated that the action is well-behaved; an obvious
extension of this work would be to go on to construct a boundary
stress tensor $\langle T_{\alpha\beta}(x^+,x^-,\theta^a) \rangle$, as was
done for the asymptotically flat case in~\cite{Mann:2005yr} and for
the linear dilaton case in~\cite{Marolf:2007ys}. This could then be
used to calculate conserved quantities. The fact that different
components of $g^{(1)}$ fall off at different rates at large $r$ may
lead to some interesting subtleties in extending the previous work to
this case; perhaps, as in the asymptotically flat case, there will be
more than one stress tensor, associated with different orders in the
asymptotic expansion. 

A central motivation for work in this direction is to better
understand holography for the plane wave. In~\cite{Marolf:2006bk}, it
was argued that a holographic dual of asymptotically flat space could
be constructed on the hyperbola at spatial infinity, calculating
two-point functions in the holographic dual from variations of the
action. It is possible that similar ideas could be applied in this
case, but there is no obvious connection between this notion of
holography and the known example. String theory on the plane wave
obtained from the Penrose limit of AdS$_5 \times S^5$ is dual to a
quantum mechanics, so it has observables depending on a single
coordinate, whereas if we were to construct a boundary stress tensor
$\langle T_{\alpha\beta}(x^+,x^-,\theta^a) \rangle$ or two-point
functions on the boundary at large $r$ from our action, we would
expect them to generically depend on all the boundary coordinates.
Our remarks in section~\ref{conf} on the relation between our notion of
asymptotically plane wave and the conformal boundary of the maximally
supersymmetric plane wave suggest that the boundary at large $r$ we
have focused on is not, at least, the whole story. To understand the
relation to holography, we probably need to study the boundaries at
constant $x^-$ in more detail, and the information coming just from
large $r$ may be misleading.

This asymptotically plane wave example thus seems to have some
interesting differences compared to previous attempts to study
holography for more general spacetimes, and we hope this work will
shed some useful light on the relation between the bulk action and the
holographic dual theory for other spacetimes, which in general remains
to be worked out.

\section*{Acknowledgements}
We are grateful for useful conversations with Veronika Hubeny, Mukund
Rangamani, Don Marolf and Amitabh Virmani. SFR was supported in part
by the EPSRC, and JL was supported by the STFC.

\providecommand{\href}[2]{#2}\begingroup\raggedright\endgroup

\end{document}